\documentclass[twocolumn,showpacs,aps,prl]{revtex4-1}

\usepackage{amsmath,amsfonts,amssymb}
\usepackage{wrapfig}
\usepackage{graphicx}
\usepackage{bbm}
\usepackage{color}

\begin{document}

\title{Verifying Genuine High-order Entanglement}\date{\today}

\author{Che-Ming Li$^{1,2}$}
\author{Kai Chen$^{3}$}
\author{Andreas Reingruber$^{2}$}
\author{Yueh-Nan Chen$^{1}$}
\email{yuehnan@mail.ncku.edu.tw}
\author{Jian-Wei Pan$^{2,3}$}

\affiliation{$^1$Department of Physics and National Center for Theoretical Sciences, National Cheng Kung University, Tainan 701, Taiwan}
\affiliation{$^2$Physikalisches Institut, Universit\"{a}t Heidelberg, Philosophenweg 12,
D-69120 Heidelberg, Germany}
\affiliation{$^3$Hefei National Laboratory for Physical Sciences at Microscale and Department of Modern Physics, University of Science and Technology of China, Hefei, Anhui 230026, China}

\begin{abstract}
High-order entanglement embedded in multipartite multilevel quantum systems (qudits) with many degrees of freedom (DOFs) plays an important role in quantum foundation and quantum engineering. Verifying high-order entanglement without the restriction of system complexity is a critical need in any experiments on general entanglement. Here, we introduce a scheme to efficiently detect genuine high-order entanglement, such as states close to genuine qudit Bell, Greenberger-Horne-Zeilinger, and cluster states as well as multilevel multi-DOF hyperentanglement. All of them can be identified with two local measurement settings per DOF regardless of the qudit or DOF number. The proposed verifications together with further utilities such as fidelity estimation could pave the way for experiments by reducing dramatically the measurement overhead.
\end{abstract}


\maketitle

\textit{Introduction.---}Entanglement is a remarkable property of multiparty quantum systems. So far, a lot of effort has been given to investigate two-level entanglement. However, general quantum states could be multilevel, meaning that their entanglement can be more complicated but possess a high potential for quantum applications. 

High-order entanglement is distinct from two-level entanglement. It offers a new insight into quantum mechanics by violating generic classical constraints \cite{dGHZ,Son2006,Collins2002}. When qudits are entangled in more than one degree of freedom (DOF) \cite{Kwiat1997,Kwiat2005}, hyperentanglement (HE) is less affected by decoherence than single-DOF multipartite entanglement \cite{Kai2007}. As resources for quantum technology, novel quantum strategies have shed light on high-performance and superior applications in quantum information \cite{Sun2003,White2004,Zeilinger2005,Zeilinger2006,Looi2008,Cabello2002, Kai2007,Barreiro2008, Lanyon2009} and quantum metrology \cite{Lloyd2008}. In addition, more and more high-order entangled systems are being created and manipulated coherently in various implementations, e.g., high-order entangled photons \cite{White2004,Zeilinger2005,Zeilinger2006,oam,Kozuma2009,Taguchi2009,Kwiat2005}. Efforts have also been devoted to create reliable superconducting qudits \cite{Neeley2009} and scalable Greenberger-Horne-Zeilinger (GHZ) states \cite{Vallone2009}.

Identifying high-order entanglement is crucial to both entanglement theories and experiments \cite{Kai2007,Cabello2002,Sun2003,Vallone2009,White2004,Barreiro2008,Looi2008, Kozuma2009,Lloyd2008,Taguchi2009,oam,Kwiat2005,Zeilinger2005,Zeilinger2006}. There has been important progress in detecting two-level entanglement \cite{Bourennane2004,Toth,Guhne2009,Vallone2008}. However, how to identify genuine multiqudit entanglement and multilevel HE remains a crucial challenge. Genuine multipartite entanglement can be generated only when all the qudits strictly participate in the correlation creation \cite{Huber2010}. On the other hand, entanglement is truly multilevel if all the levels of qudit are involved in the quantum correlation \cite{Sanpera2001}. Although violations of classical constraint could reveal entanglement \cite{dGHZ,Son2006, Collins2002}, it can not detect the genuineness of multilevel and multipartite nature. To detect genuine multipartite entanglement, a general criterion has been proposed recently by measuring identical two copies of multilevel systems \cite{Huber2010} with careful experimental design \cite{Walborn 2006}. While the criteria \cite{Sanpera2001} assisted by measuring quantum state fidelity could provide information about truly multilevel entanglement, the required experimental resources in terms of local measurement setting (LMS) increases exponentially with the qudit number in each DOF \cite{Thew2002}. There still lacks a unified and efficient method for verifying genuine multipartite multilevel entanglement and truly multilevel HE. 

In this Letter, we will propose first a novel multilevel-dependent criterion for high-order entanglement identification. Afterwards, with the criterion, we will show that states close to genuine multilevel multipartite graph states, which constitute a large and highly significant class of multipartite entangled states in physics \cite{Graph}, can be efficiently identified regardless of the number of particles. In addition, we will provide a solution in all the related experiments necessary for detecting genuine multilevel HE \cite{Kwiat2005}. The first fidelity estimation of high-order entanglement without complete fidelity measurements will also be given \cite{si}. The entanglement verification and fidelity estimation can be readily applied to the present experiments \cite{White2004,Zeilinger2005,Zeilinger2006,oam,Kozuma2009,Taguchi2009,Kwiat2005,Neeley2009,Vallone2009}. More instances for showing the power of the criterion such as verifiying multiqudit supersinglets \cite{Cabello2002} and constructing generic multilevel multipartite Bell inequalities will also be presented \cite{si}.  

\begin{figure}[t]
\includegraphics[width=3.6 cm,angle=0]{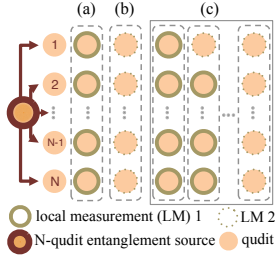}
\caption{(Color online). Each measurement in a LMS is chosen by the corresponding qudit holder for entanglement verification. All the $N$ qudits are measured locally by (a) the first measurement, (b) the second local measurements. (a) and (b) together are used to detect $N$-qudit GHZ states with the criterion (3). (c) All the $2^{N}$ LMSs for $N$-partite states are typically required by the kernel of Bell-type inequality \cite{Son2006}.}
\label{fig:epsart}
\end{figure}

Our strategy for investigating correlations between qudits highlights the statistical dependence between outcomes of measurements performed on $d$-level $N$-partite systems. We assume that two possible measurements per DOF can be performed on each spatially seperated particle and that each local measurement has $d$ possible outcomes: $v\in 
\textbf{v}=\{0,1,...,d-1\}$. As illustrated in Fig. 1, all the $N$ particles are assumed to be locally measured in parallel with a LMS. After sufficient runs of such measurements have been made, a joint probability distribution for any two subsystems could be derived from the experimental outcomes. If one of the subsystems, $A$, is composed of $a$ particles and the other subsystem, $B$, is composed of $b$ particles, then there are $d^{a}$ for $A$ and $d^{b}$ for $B$ possible measurement outcomes, $\text{v}_{a,m}$ and $\text{v}_{b,n}$, for $m=0,1,...,d^{a}-1$ and $n=0,1,...,d^{b}-1$.

\textit{Multilevel-dependent criterion.---}Independence identification is an important method for examining correlation between outcomes \cite{Freedman2007}. Two events $\text{v}_{a,m}$ and $\text{v}_{b,n}$ are statistically independent if and only if the joint probability $p(\text{v}_{a,m},\text{v}_{b,n})$ is a product of two individual marginal probabilities: $p(\text{v}_{a,m})p(\text{v}_{b,n})$. In order to extend statistical independence to a multilevel-dependent criterion for entanglement verification, we propose complex polynomials of the form: 
\begin{equation}
R_{n}\mathrel{\mathop:}=\sum_{m=0}^{d^{a}-1}c_{mn}\exp (i\phi _{mn})p(\text{v}_{a,m},\text{v}_{b,n}),
\end{equation}
for $n=0,1,...,d^{b}-1$. The phases $\phi _{mn}$'s are designed to satisfy the constraint: $\sum_{n=0}^{d^{b}-1}c_{mn}\exp (i\phi _{mn})=0$, where $c_{mn}\in \{0,1\}$. The multilevel-dependent criterion is then described by the statement: If measurement results show that every $R_{n}$
posses the same phase, then the outcomes of measurements for the two subsystems are statistically dependent.

The proof of the statement is straightforward. If the two subsystems $A$ and $B$ are independent with respect to the events of measurement, according to the definition of statistical independence we have $R_{n}=S_{n}p(\text{v}_{b,n})$, where $S_{n}=\sum_{m=0}^{d^{a}-1}c_{mn}\exp (i\phi _{mn})p(\text{v}_{a,m})$. Since $\sum_{n}S_{n}=0$, every $R_{n}$ should not have the same
complex argument, whereas a contradiction reveals the dependence between
the measurement results.

\begin{figure}[t]
\includegraphics[width=5.5 cm,angle=0]{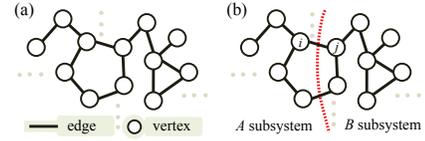}
\caption{(Color online). Qudit graph states. (a) A genuine $N$-qudit graph state can be represented by a fully-connected graph $G(V,E)$ \cite{Looi2008,Graph}. The graph $G$ consists of the set $V$ of vertices with cardinality $|V|=N$, representing the qudits, and the set $E$ of edges each of which joins two vertices, representing interacting pairs of qudits of the graph state. An edge, $(i,j)\in E$, corresponds to an unitary two-qudit transformation among the two qudits (vertices) $i$ and $j$ by $U_{(i,j)}\!\!=\!\!\sum_{v=0}^{d-1}\left|v\right\rangle_{ii}\!\left\langle v\right|\otimes (Z_{j})^{v}$, where $\{\left|v\right\rangle_{i}\}$ is an orthonormal basis of the $i$th qudit and $Z_{j}=\sum_{k=0}^{d-1}\omega^{k}\left|k\right\rangle_{jj}\!\left\langle k\right|$, $\omega=\exp(i2\pi/d)$. An explicit state vector of a given graph $G$ is generated by applying $U_{(i,j)}$ to an initial state $\left|f_{0}\right\rangle=\bigotimes_{k=1}^{N}F_{k}\left|0\right\rangle_{k}$\cite{Sun2003}: $\left|G\right\rangle=\prod_{(i,j)\in E}U_{(i,j)}\left|f_{0}\right\rangle$, where $F_{k}$ is the quantum Fourier transformation defined by $F_{k}\left|v'\right\rangle_{k}=\sum_{v=0}^{d-1}\omega^{v'v}\left|v\right\rangle_{k}/\sqrt{d}$. (b) The state $\left|G\right\rangle$ can be decomposed in the Schmidt form with respect to a fixed bipartite splitting as illustrated by the red-dashed line \cite{si}. Note that there always exists at least a pair $(i,j)\in E$ where each vertex belongs to the opposite subsystem. Since there are $d$ nonvanishing terms in the Schmidt form, the Schmidt rank of the graph state is $d$. One can characterize mixed states in the same instructive way as Schmidt rank for pure states. The Schmidt number ($s$) of a mixed state \cite{Terhal2000} describes the minimum Schmidt rank of the pure states that are required to construct the density matrix.}
\label{fig:epsart}
\end{figure}

The property that all $R_{n}$ are not parallel is not only a
characteristic feature of two independent subsystems, but also
a basis for a quantitative analysis of statistical dependence. Let us consider the quantity $|\sum_{n}R_{n}|$ for independent
systems. Since the variety  of different phases, it is clear that if $p(\text{v}_{a,m})<1$ and $p(\text{v}_{b,n})<1$ for all events, then $|\sum_{n}R_{n}|=|\sum_{n}S_{n}p(\text{v}_{b,n})|<1$, while $|\sum_{n}S_{n}p(\text{v}_{b,n})|=1$ exists only when $p(\text{v}_{a,m'})=p(\text{v}_{b,n'})=1$ for some $m'$ and $n'$, which can be strictly identified by obtaining the outcomes $\text{v}_{a,m'}$ and $\text{v}_{b,n'}$ with certainty. This quantitative discrimination between
statistical dependence and independence is of the essence to signify truly
multipartite multilevel entanglement.

\textit{Detecting genuine many-qudit graph states.---}In order to examine whether an experimental output state, $\rho$, is a genuine high-order entangled state close to a graph state $\left|G\right\rangle$ [see Fig.~2(a)], the first step in our strategy is to specify the correlations between qudits. We feature the most typical correlation quality of $\left|G\right\rangle$ by a sum of all related local operators for multilevel-dependent criterion. The correlation between the $k$th qudit and all qudits of its neighborhood is described by
\begin{equation}
\hat{a}_{k}\mathrel{\mathop:}=\frac{t_{d}}{d-1}\bigg[d\!\!\!\!\!\sum_{\scriptscriptstyle{{\overset{\scriptscriptstyle{v,v_{\!i}}}{v+\sum_{i}\!\!v_{\!i}\doteq 0}}}}\!\!\!\!\!\big(F_{k}^{\dag}\left|v\right\rangle_{\!kk\!\!}\left\langle v\right|F_{k}\!\bigotimes_{i\in N(k)}\!\left|v_{\!i}\right\rangle_{\!ii\!\!}\left\langle v_{\!i}\right|\big)\!-\!\hat{\text{I}}\bigg],
\end{equation}
where $\doteq$ denotes equality modulo $d$, $N(k)$ is the set of all $i$'s satisfying $(k,i)\in E$, and $\hat{\text{I}}$ denotes the identity operator. As has been shown \cite{si}, $\hat{a}_{k}$ is constructed according to the multilevel-dependent criterion, and from which one can see that $t_{d}$ is the value associated with the map of measurement events: $t_{2}=1$, $t_{3}=2$, and $t_{d}=(d-1)(t_{d-2}+t_{d-1})$ for $d\geq 4$. See Figs.~3(a),(b).
  
\begin{figure}[t]
\includegraphics[width=7.7 cm,angle=0]{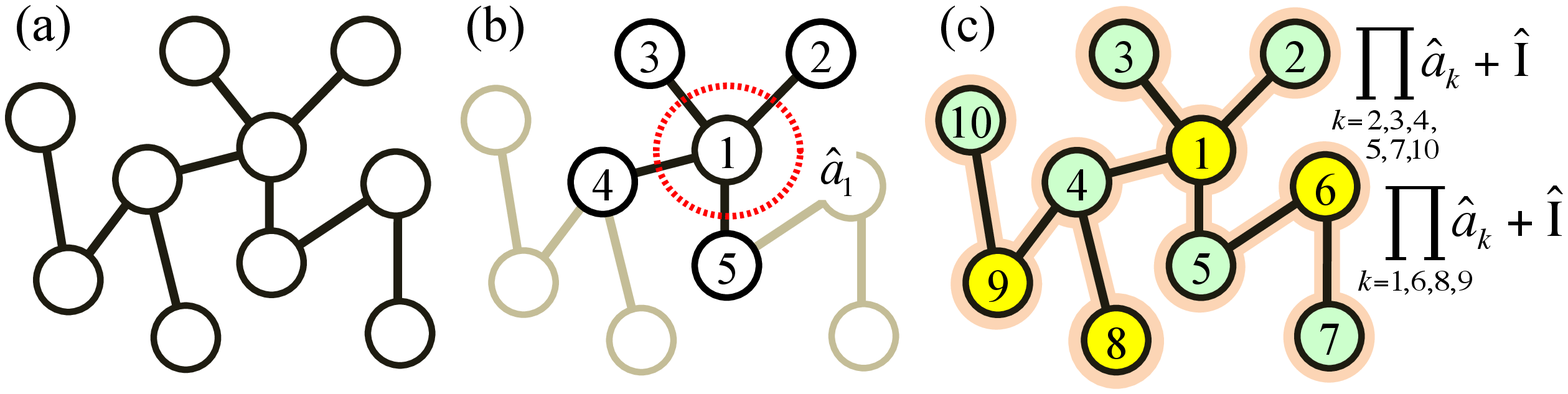}
\caption{(Color online). Construction of the kernel of entanglement criterion. For a $10$-qudit state with the graph (a), the kernel of the entanglement criterion (3) is constructed in two steps: (b) Signify the correlation between each qudit and all its adjacent qudits in the operator $\hat{a}_{k}$ $(2)$. Here, the correlation feature of the first qudit is highlighted by red-dashed circle. (c) Combine $\hat{a}_{k}$'s according to the coloring of the graph. If the vertices of the graph $G$ can be divided into $q$ sets, say $Y_{j}$ for $j=1,2,\ldots,q$, and the vertices of each set are given a color such that adjacent vertices have different colors, then the graph is called a $q$-colorable graph \cite{Graph}.}
\label{fig:epsart}
\end{figure} 
   
The second step is to combine all $\hat{a}_{k}$'s according the graph coloring and then construct the kernel of entanglement criterion [see Fig.~3(c)]. We posit that if $\rho$ shows that
\begin{equation}
\sum_{j=1}^{q}\left\langle\prod_{k\in Y_{j}}\!\!\frac{\hat{a}_{k}+\hat{\text{I}}}{t_{d}+1}\right\rangle>\frac{1}{d}(l-1)(q-\eta_{q})+\eta_{q},
\end{equation}
then $\rho$ is a genuine $N$-partite entangled state with a Schmidt number $s\geq l$, where $l\in\{2,3,...,d\}$ and $\eta_{q}=(q-1)+(1-\frac{t_{d}}{d-1})^{f_{q}}(1+t_{d})^{-f_{q}}$. $f_{q}=2$ for any graphs having at least one set of color $Y_{k}$ with $|Y_{k}|>1$ and $f_{q}=1$ otherwise. For $l=2$, the criterion (3) detects genuine $N$-partite entanglement with $s\geq 2$. To detect entangled states having higher Schmidt numbers, higher $l$ should be chosen for the criteria. Since a $d$-level graph state has $s=d$ [Fig. 2(b)], states detected by a criterion with $l=d$ are then genuinely $d$-level in the most proximity of $\left|G\right\rangle$. See the Appendix for the proof of the criterion.

\begin{figure}
\includegraphics[width=5.8 cm,angle=0]{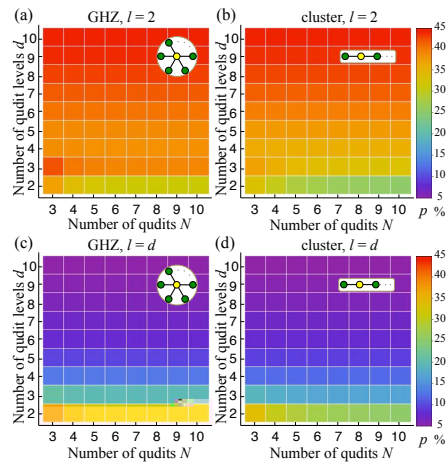}
\caption{(Color online). Noise tolerance of entanglement criterion. For $l=2$, the noise tolerance is up to $p_{\text{noise}}<1/q$ for large $d$, independent of the number of qudits. Since Bell, GHZ, and cluster states correspond to two-colorable bar, star, and chain graphs, respectively, the criteria are very robust. For Bell states, our criterion tolerates mixing with white noise if $p_{\text{noise}}<50\%$, independent of $d$. For (a) GHZ, and (b) cluster states, up to $50\%$ noise is tolerated for any $N$. For $l=d$, the robustness of the criterion is mainly affected by $d$. The robustness degradation with increasing qudit number is small, as depicted in (c) and (d) for GHZ and cluster states, respectively.}
\label{fig:epsart}
\end{figure}

Two local observables are measured herein: one has the nondegenerate eigenvectors $\{\left|v\right\rangle_{k}\}$ and the other has the eigenvectors $\{F^{\dag}_{k}\left|v\right\rangle_{k}\}$. Then $q$ LMSs are sufficient to implement the verification. It is also worth noting that the proposed criterion is robust against noise as shown in Fig. 4. In the presence of white noise the pure state $\left|G\right\rangle$ becomes $\rho(p_{\text{noise}})=p_{\text{noise}}\hat{\text{I}}/d^{N}+(1-p_{\text{noise}})\left|G\right\rangle\!\!\left\langle G\right|$, where $p_{\text{noise}}$ is the probability of uncolored noise. The state $\rho(p_{\text{noise}})$ is certainly identified as true $N$-multipartite entanglement with $s\geq l$ if $p_{\text{noise}}<p$, where $p=(q\!-\!\eta_{q})(d-l+1)/\{d\big[q\!-\!\sum_{k}(t_{d}+1)^{-|Y_{k}|}\big]\}$ \cite{si}.

\textit{Comparison.---}One can detect genuine multiqudit entanglement with $s\geq l$ by the projector-based criterion: $\left\langle\left|G\right\rangle\!\!\left\langle G\right|\right\rangle>\frac{1}{d}(l-1)$ (see Refs. \cite{Sanpera2001,si,Bourennane2004}). While the resistance of this criterion against noise is up to $2$
times higher than the proposed criterion \cite{si}, however, it is difficult to realize it experimentally. Decomposing the projector $\left|G\right\rangle\!\!\left\langle G\right|$ into a sum of \emph{locally measurable operators} is necessary in this method. For decomposition of $\left|G\right\rangle\!\!\left\langle G\right|$, $2(d^{N}-1)$ settings \cite{Thew2002} are heavily needed for further detection tasks. One can alternatively make a quantum state tomography for the multiqudit states and then invoke the projector-based criterion. Nevertheless, huge $d^{2N}-1$ LMSs \cite{Thew2002} are necessary for the tomographic analysis. For the criterion by Huber et al. \cite{Huber2010} to verify genuine multipartite entanglement, one needs identical two copies of states for each round of measurement and requires $2N(N-1)$ LMSs for the most efficient detection in the illustrated criteria. Compared with these approaches, our scheme is rather efficient: \emph{two} LMSs are sufficient for verifying genuine Bell, GHZ, cluster and any two-colarable graph states [Figs.~1(a) and 1(b)]. Furthermore, our criterion can be used to estimate the quality of the prepared state without full fidelity measurements \cite{si}. 

\textit{Detecting genuine multilevel HE.---}We continue to illustrate the scheme with verification of existing multilevel $N$-DOF HE \cite{Kwiat2005}: $\left| H\right\rangle =\bigotimes_{k=1}^{N}\frac{1}{d_{k}}\sum_{v=0}^{d_{k}-1}\sum_{v'=0}^{d_{k}-1}\omega^{vv'}\left| v\right\rangle_{\!A_{k}}\otimes\left|
v'\right\rangle_{\!B_{k}}$, where $A_{k}$ $(B_{k})$ denotes the $k$th DOF of the subsystem $A$ $(B)$ with $d_{k}$ levels. The entangled state in each DOF is a two-qudit graph state. To identify genuine HE, it is crucial to recognize the difference
between the state $\left| H\right\rangle $ and a state with biseparable
structure in the hyperentangled sense \cite{Vallone2008}: $\left| h_{b}\right\rangle =\left|
h_{1}\right\rangle _{A_{k}b_{1}}\otimes \left| h_{2}\right\rangle
_{B_{k}b_{2}}$, where $\left|h_{1}\right\rangle_{A_{k}b_{1}}=\sum_{i=0}^{d_{k}-1}\!\!c_{ib_{1}}\left|
i\right\rangle _{\!A_{k}}\otimes\left| u_{i}\right\rangle _{b_{1}}$ and $\left|h_{2}\right\rangle_{B_{k}b_{2}}\!\!=\!\!\sum_{i=0}^{d_{k}-1}\!\!c_{ib_{2}}\left|
i\right\rangle _{\!B_{k}}\otimes\left| u_{i}\right\rangle _{b_{2}}$. $\{A_{k},b_{1}\}$ and $\{B_{k},b_{2}\}$ constitute the
set of all DOFs, where the sets $b_{1}$ and $b_{2}$ are disjoint. The following criterion is introduced to distinguish genuine HE from correlations  mimicked by biseparable states: $\left\langle\!\!\bigotimes_{k=1}^{N}\!\!\frac{\hat{a}_{1}^{(k)}+\hat{a}_{2}^{(k)}+t_{d_{k}}\hat{\text{I}}}{3t_{d_{k}}}\!\right\rangle>1-\frac{D(d-1)}{3d(D-1)}$, where $\hat{a}_{1}^{(k)}$, $\hat{a}_{2}^{(k)}$ are of the form $(2)$ for the $k$th DOF, $d=\min\{d_{k}\}$, and $D=\max\{d_{k}\}$. Refer to Appendix for the proof. This criterion can be efficiently implemented because each DOF needs only two LMSs. The measurement results also can be used to examine multilevel genuineness in each DOF by $(3)$. A state mixing with white noise $\rho(p_{\text{noise}})$ is detected as state with genuine HE if $p_{\text{noise}}<\frac{(1-\frac{1}{d})}{3(1-\frac{1}{D})(1-3^{-N})}$. E.g., for $D=d$, the noise tolerance is at least $33\%$ for any number of DOFs.

Our work is partially supported by the NSC, Taiwan, under Grant No. 99-2112-M-006-003 and No. 98-2112-M-006-002-MY3. This work is also supported by the European Commission through the ERC Grant, the CAS, the National Fundamental Research Program of China under Grant No. 2011CB921300,
and the NNSFC.

\textit{Summary and outlook.---}We have introduced a general criterion of multilevel dependence to demonstrate an efficient verification of high-order entanglement regardless of the number of particles and DOFs. The importance of our scheme are three fold. First, the detection scenario is applicable to any Bell-type experiment \cite{Son2006,Collins2002} and quantum protocols \cite{Cabello2002,Sun2003,White2004,Zeilinger2005,Zeilinger2006,Looi2008,Lloyd2008}. Second \cite{si}, our criterion can be readily used to speed up the existing experimental verification \cite{White2004,Zeilinger2005,Zeilinger2006,oam,Kozuma2009,Taguchi2009,Neeley2009,Vallone2009}, and to provide a \emph{quantitative} estimation of experimental state fidelity. Third, the present detection offers a scalable detection method in experiments. In addition, the idea presented in this work opens up some questions: for instance, how one can detect genuine high-order entanglement of general \emph{mixed systems} with the proposed criterion, or how one can construct generic Bell inequalities for general qudit graph states.

\textit{Appendix.---}We will show that all states identified by the proposed criterion $(3)$ are also detected by the projector-based criterion \cite{Lewenstein2001}, which is equivalent to show that $w_{G}-\gamma_{G} w'_{G}\geq 0$. $\gamma_{G}$ is some positive constant, $w'_{G}\mathrel{\mathop:}=\frac{1}{d}(l-1)\hat{\text{I}}-\left|G\right\rangle\!\!\left\langle G\right|$ for the projector-based criterion, and $w_{G}\mathrel{\mathop:}=[\frac{1}{d}(l-1)(q-\eta_{q})+\eta_{q}]\hat{\text{I}}-\sum_{j=1}^{q}\prod_{k\in Y_{j}}\!\!\frac{\hat{a}_{k}+\hat{\text{I}}}{t_{d}+1}$ for the proposed criterion. Since the operator $w_{G}$ as well as $w'_{G}$ are diagonal in the graph state basis, $w_{G}-\gamma_{G}w'_{G}$ is also diagonal \cite{si}. Here a graph state basis is composed of $d^{N}$ orthonormal states $\left|G_{k}\right\rangle=\prod_{(i,j)\in E}U_{(i,j)}\left|f_{k}\right\rangle$ for $k=0,1,...,d^{N}-1$. $\{\left|f_{i}\right\rangle=\bigotimes_{k=1}^{N}F_{k}\left|i\right\rangle|i=0,1,...,d^{N}-1\}$ is an orthonormal basis, where $\left|0\right\rangle=\left|0\right\rangle_{1}...\left|0\right\rangle_{N}$, ..., $\left|d^{N}-1\right\rangle=\left|1\right\rangle_{1}...\left|1\right\rangle_{N}$. When $\gamma_{G}=q-\eta_{q}$ the diagonal elements are then all non-negative \cite{si}. Similarly, by comparison with a projector-based criterion for genuine HE: $\left\langle\left|H\right\rangle\!\!\left\langle H\right|\right\rangle>\frac{1}{d}$, all states detected by our criterion are also identified by the projector-based criterion. The bound $\frac{1}{d}$ directly follows from the derivation: $\left\langle\left|H\right\rangle\!\!\left\langle H\right|\right\rangle>\max_{\left| h_{b}\right\rangle \in \mathbf{b}}|\left\langle h_{b}\right| \!H\rangle |^{2}\geq\max_{d_{k}}\frac{1}{d_{k}}=\frac{1}{d}$, where $\mathbf{b}$ denotes the set
of biseparable states in the hyperentangled sense. Since $w_{H}-\gamma_{H} w'_{H}\geq 0$, where $\gamma_{H}=\frac{D}{3d(D-1)}$, $w_{H}$ and $w'_{H}$ denote the operators corresponding to the proposed criterion and the projector-based criterion, respectively, we prove that our criterion can be used to identify genuine HE.

\end{document}